\documentclass[12pt]{article}
\begin{document}

\title{Fermion Masses from SO(10) Hermitian Matrices}
\author{
R.~G.~Moorhouse\\[.4cm]
\small Department of Physics and Astronomy\\
\small University of Glasgow,Glasgow G12 8QQ, U.K.
\thanks{email: g.moorhouse@physics.gla.ac.uk}\\}
\maketitle
\begin{abstract}
\noindent

 Masses of fermions in the SO(10) 16-plet are constructed
 using only the ${10},{120}$ and $\overline{126}$ scalar multiplets.
 The mass matrices are restricted to be hermitian and the theory
 is constructed to have certain assumed quark masses,charged
 lepton masses and CKM matrix in accord with data. The remaining
 free parameters are found by fitting to light neutrino masses
 and MSN matrices result as predictions.
\end{abstract}

\section{Introduction}
The simplest SO(10) treatment of fermion masses, with the fermions
being in SO(10) 16-plets, uses the composition of 256 pairings of
fermions into the ${10},{120}$ and $\overline{126}$ SO(10)
 representations.There 
are 3 different 16-plets for the 3 generations of 
fermions and the fermion pairs are coupled to  scalar 
(generalized Higgs) bosons
 in the same SO(10) representations with $3 \times 3$ Yukawa
coupling matrices; and when the scalar bosons develop vevs the
 $3 \times 3$ fermion mass matrices are generated. The philosophy
 in this paper is to assume fermion masses only arise through 
 coupling into these representations.

In the coupling to the $\{10\}$ and $\{\overline{126}\}$  bosons 
 SO(10) gives equal
coupling to the mass terms $\bar{\psi}_A\psi_B$ and $\bar{\psi}_B\psi_A$
 where  $A,B$ label different 16-plets of fermions. Here they are taken 
to be generation labels and thus the Yukawa coupling matrices can 
be taken to be symmetric without loss of generality; contrairiwise
for the $\{120\}$  the coupling of the mass terms
 $\bar{\psi}_A\psi_B$ and $\bar{\psi}_B\psi_A$ are of equal magnitude
 but opposite signs so that the $\{120\}$ Yukawas can be taken 
anti-symmetric \cite{MS}. For this, chiral, theory there is no constraint, in 
principle, that the Yukawa or mass matrix elements be real.

In the very many SO(10) fermion mass investigations, ranging from 
simple hypotheses to complicated varieties of SUSY GUTS, there has 
been until recently a preponderance of hypotheses with no $\{120\}$ 
Higgs particles \cite{GMN,FO,BM,BSV,DMM1}. This rather arbitrary neglect 
seems to have been 
motivated, not unnaturally, by the search for simplicity,solvability and
predictive power. Latterly, partly influenced by neutrino data, there have
been more papers taking account of the $\{120\}$
 \cite{Bertolini,GK1,LGK} - some of these still giving a predominant role
 to $\{10\}$ and $\{126\}$ {\it ab initio}. This present 
paper is an example of the $\{120\}$ filling a necessary role, that of
supplying the CP non-invariant, imaginary part, of the CKM matrix.
 
  There have been continual developments of SO(10) type theories - among 
 other reasons there have been the demands of data matching. Begining
 with theories having real Yukawa coefficients and real vacuum expectation
 values these have ranged  to much more sophisticated models such as many
 having vevs with arbitrary phases to be determined from data extrapolated
 to a GUT scale (as for example in refs. \cite{GMN,GK1}).
 In this paper a simplification from general possibilities is chosen - 
 that is the contribution to the fermion mass matrices from the $\{10\}$
 and $\{\overline{126}\}$  to be real and symmetric while that from the
 $\{120\}$ be pure imaginary and anti-symmetric, resulting in Hermitian mass 
 matrices. That condition on the $\{10\}$ and $\{\overline{126}\}$
 is familiar,
especially in early papers, and can be presented as 'naturally' implemented
using real Yukawas; in the formalism given in the next section the above 
condition on the $\{120\}$ contribution can be presented as equally
'natural'; that is the mathematical formalism, as outlined in section 3,
 suggests the hermiticity as the simplest choice.

 There is also an important physical motivation for the choice of
 hermiticity, this being the association in the formalism with parity 
 invariance at high energy. If the $\gamma_5$ terms are absent in the
 mass terms these are straightforwardly parity invariant, giving a
 limiting case of the chiral theory for the quark and charged lepton 
 equations. And the absence of the $\gamma_5$ terms gives the hermiticity 
  and then also the formalism yields the mass contributions of the $\{10\}$ 
 and $\{126\}$ as CP conserving and that of the $\{120\}$ as CP violating.
 It can thus be said that there exists at least one basis (implicit in the
 formalism) in which the above mass matrices are hermitian. The one
 exception to high energy parity invariance and hermiticity 
 in the mass terms of the model results indirectly, that is 
 in the light neutrino masses from the contribution of the
  original seesaw (often known as 'Type I') in which the product of three
 hermitian mass matrices yields a non-hermitian matrix. The details for 
 the model are in section IV below and the Appendix.
 (As has been emphasized recently \cite{ZAJM} parity
 restoration in proceeding to high energies was one of the main 
 motivations for the left-right symmetry principle which features in the 
 next section.)

 The hermiticity, as derived below, is more precisely that of  
 $3 \times 3$ flavour matrices associated with the symmetry representations,
 linear superpositions of which form the hermitian mass matrices of
 the quarks and charged leptons - and one other contribution to
 the light neutrino mass matrix

 At least two papers \cite{DMM2,GK2} incorporating the ${120}$ along
 with hermitian mass matrices have previously been published. Both of
 these are set in MSSM theory and one in particular \cite{DMM2} has much
 detailed discussion of the various symmetry breakings in the MSSM
 context. The present paper is based on SO(10) but has no committment to
 supersymmetry or details of a Higgs mechanism or other higher theory.

The experimental data input to this model are the quark and charged lepton
masses and the CKM matrix; so the parameters of the model have to be chosen 
to accomodate these numbers some of which carry considerable uncertainty;
 there can be further uncetainties from extrapolation to higher energies.
The vital question then is can one adjust the very few remaining
{\it completely}
 unknown parameters so that the model be compatible with the likely masses
 and MSN matrix of the three light neutrinos. In this paper the SO(10) 
vevs (or substitute mechanism) giving rise to the masses are parameters
to be chosen to match data and no particular  Higgs potential (or other 
mechanism) is postulated..  
 
 In section II the SO(10) Clifford algebra formalism for the fermion masses
 is given. Section III outlines the calculation with the Hermitian matrix
 hypothesis and gives the resulting mass formulae in terms of Yukawa
 coefficients and scalar vacuum expectation values; while in section IV 
 the assumptions of the theory allow the expression of the mass matrices 
 in terms of Hermitian matrices and real ratios of vevs and also allow
 the incorporation of the quark and charged lepton mass data and 
 CKM complex matrix data. Section V deals with the neutrino masses with 
 emphasis on  simple examples. One illustrates  tri-bimaximal mixing arising
 from neutrino masses suggested by the experimental data. 

\section{SO(10) and its 16-plets}

 The Clifford algebra formalism of SO(10) \cite{MS} is based 
 on ten gamma matrices $(\Gamma_1,....,\Gamma_{10})$ giving the 45 
 generators
 $\Sigma_{\mu\nu} \equiv [\Gamma_{\mu},\Gamma_{\nu}]/2i$.
 The gammas can also be expressed through the creation
 and annihilation operators $\chi_j,\chi_j^{\dagger},(j=1,...5)$ where
 $$ \Gamma_{2j-1} =-i( \chi_j-\chi_j^{\dagger}), 
 \Gamma_{2j} =( \chi_j+\chi_j^{\dagger})$$.

 The fermion 16-plet,$\psi_+$ of positive 10d chirality is expressed as 
\begin{equation}
\left|\psi_+\right> \equiv \left|0\right>\psi_0 +
{1 \over 2}\chi_j^{\dagger}\chi_k^{\dagger}\left|0\right>
\psi_{jk} + {1 \over 24}\epsilon^{jklmn} 
\chi_k^{\dagger}\chi_l^{\dagger}\chi_m^{\dagger}\chi_n^{\dagger}
\left|0\right>\tilde \psi_j
\label{21}
\end{equation}
where $\psi_0, \psi_{jk}, \tilde \psi_j$ are 2-component left-handed
 spinors of the particular generation. The assignment to the leptons 
and the quark $SU(3)_{colour}$ triplets is, in an obvious notation
with $a,b$ being indices $1,2,3$ and $d,u$ being colour triplets:
$\tilde \psi_a \approx (d_R)^c$;$  \psi_{a5} \approx d_L $;
 $ \psi_{ab} \approx (u_R)^c$; $ \psi_{a4} \approx u_L $;
$\tilde \psi_4 \approx e^-_L;$;$ \psi_{45} \approx (e^-_R)^c$;
$\tilde \psi_5 \approx n_L$ ;$ \psi_0 \approx -(n_R)^c$ 
 where $n_R$ denotes right-handed (heavy) neutrinos.

The conjugate of equation (\ref{21}) transforming appropriately
 under SO(10) is 
\begin{equation}
\left< \psi_+^{\star} \right| B_{\Gamma} \equiv 
\left < \tilde 0 \right| \psi_0^T + 
\psi_{ij}^T \left < \tilde 0 \right|
 ({1 \over 2}\chi_j^{\dagger}\chi_i^{\dagger}) +
 \tilde {\psi} _j^T \left < 0 \right| \chi_j
\label{22}
\end{equation}
where $B_{\Gamma} \equiv i \Gamma_1 \Gamma_3 \Gamma_5 \Gamma_7 \Gamma_9$ 
and $\left < \tilde 0 \right| \equiv \left < 0\right| 
\chi_5\chi_4\chi_3\chi_2\chi_1$.

Let the suffices A and B be generation indices for the 3 generations.
Then, letting C be the charge conjugation matrix
\begin{equation} 
\left<A\right|X\left|B\right> \equiv 
\left< \psi_{+A}^{\star} \right| B_{\Gamma}C^{-1}X\left|\psi_{+B}\right>
\label{23}
\end{equation}
with $X=\Gamma_{\mu}$ or $X=\Gamma_{\mu}\Gamma_{\nu}\Gamma_{\rho}$ or
$X=\Gamma_{\mu}\Gamma_{\nu}\Gamma_{\rho}\Gamma_{\sigma}\Gamma_{\tau}$
form the SO(10) representations 10 or 120 or 126 respectively.
 Under the action of a symmetry group generator $\Sigma_{\mu\nu}$
\begin{equation} 
\left<A\right|X\left|B\right> \to 
 \left<A\right|[X,\Sigma_{\mu\nu}]\left|B\right>.
\label{23b}
\end{equation}
Multiplying respectively by scalar fields $\phi_{\mu}$,
$\phi_{\mu\nu\rho}$,$\phi_{\mu\nu\rho\sigma\tau}$ of the same 
representations gives SO(10) invariants, for example
$$\left<A\right|\Gamma_{\mu}\left|B\right>\phi_{\mu} \equiv 
\left< \psi_{+A}^{\star} \right| B_{\Gamma}C^{-1}
\Gamma_{\mu}\left|\psi_{+B}\right>\phi_{\mu}$$
 which for colourless  neutral
vevs contribute to the elements of the 3 by 3 generation mass 
matrices of the quarks and leptons \cite{MS}.

 It is convenient to classify these vevs as the neutral colourless
members of multiplets of the Pati-Salam subgroup of SO(10).
Selecting the generators formed by $(\Gamma_1,....,\Gamma_6)$ 
 gives the 15 generators of an SO(6) subgroup, and likewise
 $(\Gamma_7,....,\Gamma_{10})$ give the 6 generators of an SO(4)
 subgroup. These realise the $SO(6) \times SO(4)$,
 otherwise $SU(4)_c \times SU(2)_L \times SU(2)_R$,
 Pati-Salam subgroup of SO(10), the $3+3$ generators of 
 $SU(2)_L \times SU(2)_R$ being linear combinations of the 6
 generators of SO(4). In terms of the creation and annihilation
 operators $(j=1,2,3)$ can give SO(6) and $(j=4,5)$ can give SO(4).
We are concerned with transition operators X, a subset of those above, 
which form $SO(6) \times SO(4)$ 
(equivalently $SU(4)_c \times SU(2)_L \times SU(2)_R$) invariant elements 
$$\sum_X\left<A\right|X\left|B\right>\phi_x$$ by coupling to the 
scalar field $\phi_x$ which transforms in the same 
 $SU(4)_c \times SU(2)_L \times SU(2)_R$
 representation as $\left<A\right|X\left|B\right>$.
The neutral, colour singlet, subset $X_0$ of X, coupling to vevs
 $\phi_{x_0}$ and multiplied by Yukawa coupling constants 
$Y_{AB}(X_0)$  
\begin{equation}
\sum_{X_0}\left<A\right|X_0\left|B\right>\phi_{x_0}Y^{\rho}_{AB}(X_0)
\label{23c}
\end{equation}
yields contributions to 3 by 3 mass matrices. Here $\rho$, being
 10 or 120 or 126, denotes the SO(10) representation to which $X_0$
 belongs; thus the Yukawas are those appropriate to an unbroken SO(10)
 symmetry. This symmetry is subsequently broken by the mass terms.

The Table shows the subsets $X_0$ contributing to the fermion masses via
$\left<A\right|X_0\left|B\right>$.
 The left hand column shows the dimensions of the
$SU(4)_c \times SU(2)_L \times SU(2)_R$ representations to which the
 $\left<A\right|X_0\left|B\right>$ belongs.
 The right hand column gives the notation for the scalar
vevs $\langle \phi_x \rangle$ labelled also by the SO(10) representation
(10, 120 or 126) of which the $X_0$ is a member.
 For example the first pair of rows contain
 the neutral members of an $SU(2)_L \times SU(2)_R$ bi-doublet and 
 so do the second, third and fourth pairs of rows. Further comments on
 this Table as well as details of how the terms in it give rise to the
 mass matrices are given in the next Section.

\begin{table*}
\begin{center}
\begin{tabular}{|c|c|c|}\hline
$SU(4)_c \times SU(2)_L \times SU(2)_R$ & $\Gamma$ products,$X_0$ & 
scalar vev \\ \hline\hline
$\{1,2,2\}$            & $ \chi_5 $              & $v_-^{10}$ 
\\ \hline
$\{1,2,2\}$            & $ \chi_5^{\dagger}$     & $v_+^{10}$ 
\\ \hline
$\{1, 2, 2\}$    & $\Gamma_7\Gamma_8 \chi_5 $ & $v_-^{120a}$ 
\\ \hline
$\{1, 2, 2\}$ &$\Gamma_7\Gamma_8 \chi_5^{\dagger}$  & $v_+^{120a}$
\\ \hline
$\{15,2,2\}$
 & $(\Gamma_1\Gamma_2 +\Gamma_3\Gamma_4 +\Gamma_5\Gamma_6)\chi_5 $
 & $v_-^{120b}$ \\ \hline
$\{15,2,2\}$
&$(\Gamma_1\Gamma_2+\Gamma_3\Gamma_4+\Gamma_5\Gamma_6)\chi_5^{\dagger}$
 & $v_+^{120b}$ \\ \hline
$\{15, 2, 2\}$
 & $(\Gamma_1\Gamma_2 +\Gamma_3\Gamma_4 +\Gamma_5\Gamma_6)
\Gamma_7\Gamma_8\chi_5 $
 & $v_-^{126a}$ \\ \hline
$\{15, 2, 2\}$
&$(\Gamma_1\Gamma_2+\Gamma_3\Gamma_4+\Gamma_5\Gamma_6)
\Gamma_7\Gamma_8\chi_5^{\dagger}$
 & $v_+^{126a}$ \\ \hline
$\{\overline{10},3,1\} $& $\chi_1\chi_2\chi_3\chi_4\chi_5^{\dagger} $
 & $v_-^{126b}$ \\ \hline
$\{10, 1, 3\} $ & $\chi_1^{\dagger}\chi_2^{\dagger}
\chi_3^{\dagger}\chi_4^{\dagger}\chi_5^{\dagger}$ & $v_+^{126b}$ 
\\ \hline
\end{tabular}
\end{center}
\caption{SO(10) Clifford algebra operators giving rise to 
 fermion masses} 
\end{table*}

\section{Yukawa coefficients, vevs, hermiticity}

 The mass terms shown in the Table break not only the SO(10)
 symmetry but also the Pati-Salam left-right symmetry
 since the operators in the middle column do not 
 commute with all the Clifford algebra Pati-Salam generators.
 As a theory of particle masses, in the present context,they are
 only appropriate for use at high energy - say near the GUT
 energy - since the use of SO(10) Yukawas as in equation
 (\ref{23c}) is only appropriate near energies where SO(10) is 
 a good symmetry. Such masses can ordinarily only be derived 
 from experimental observation by the use of RGE equations
 such as those we shall make use of.

 The symmetry breaking is fairly clear. For example the 
last row can not only supply the heavy Majorana neutrino mass 
necessary for the Type I seesaw but also breaks the 
$SU(2)_R$ symmetry because of the triplet component. It may be noted
that the ninth row, necessary for the Type II seesaw breaks 
$SU(2)_L$ but because of the numerics of the neutrino masses
 only by a tiny amount. The combination
 $(\Gamma_1\Gamma_2+\Gamma_3\Gamma_4+\Gamma_5\Gamma_6)$ 
 occuring in rows 5,6,7,8 is proportional to the B-L generator
 and thus breaks the $SU(4)_c$ symmetry:
 $SU(4)_c \to SU(3)_c \times U(1)_{B-L} $. In addition the 
 operators in rows 9 and 10 do not commute with $(B-L)$ so a non zero
 value of either of the corresponding vevs gives
 $SU(4)_c \to SU(3)_c$.  

 Thus if all the vevs in the Table are non-zero many symmetries are 
 multiply broken.

The lines of the Table with one, three or five $\Gamma_{\mu}$ correspond
respectively to subsets of the $\{10\}$, $\{120\}$ or $\{126\}$ 
algebras of SO(10). Fixing conventions  we illustrate  by
 outlining the down quarks mass matrix calculation. 
 
 \subsection{One $\Gamma$}.
  From the $\{10\}$ of SO(10) these are the first two lines of the Table 
 providing the colour singlet
 members of an $SU(2)_L \times SU(2)_R$ bi-doublet. 
 To corespond to colour singlet, neutral vevs $X_0$ must
 be some combination of $\Gamma_9$ and $\Gamma_{10}$. The combinations
 $\chi_5$ and $\chi_5^{\dagger}$
 (being $(\Gamma_{10} \pm i\Gamma_9)/2)$ are chosen with the 
 corresponding vevs, $v_{\pm}^{10} 
 \equiv \left<(\phi_{10} \pm i\phi_{9})\right> $.
  
 Evaluation of the resulting expressions ($Y^{10}_{AB}$ being Yukawa 
 coefficients)
\begin{equation}
\left< \psi_{+A}^{\star} \right| B_{\Gamma}C^{-1}
(\chi_5 v_-^{10} + \chi_5^{\dagger} v_+^{10})
\left|\psi_{+B}\right>Y^{10}_{AB} +h.c.
\label{31}
\end{equation}
 gives directly contributions to the mass matrix elements of the quarks,
 charged leptons and neutrinos as can be seen on inspection of eqns.
 (\ref{22}),(\ref{23}) and (\ref{23b}), using the operator algebra. Consider
 for example the contribution to the {\it two} down quark mass matrix  
 elements  A,B and B,A. For one $\Gamma$ these are all included in
\begin{eqnarray}
[\left< \psi_{+A}^{\star} \right| B_{\Gamma}C^{-1}
\chi_5 \left|\psi_{+B}\right>Y^{10}_{AB} +\nonumber\\
\left< \psi_{+B}^{\star} \right| B_{\Gamma}C^{-1}
\chi_5 \left|\psi_{+A}\right>Y^{10}_{BA}]v_-^{10} +h.c.
\label{32}
\end{eqnarray}
This is expressed in terms of left, $d_L$, and right,$d_R$, Weyl 
spinors using $C^{-1}=\sigma_2$ and equations (\ref{22}),(\ref{23})
 with the correspondences
$\tilde \psi_a =\sigma_2 (d_{aR})^*$;$  \psi_{a5}= d_{aL}$, $a$ 
being the colour index. The result is

\begin{equation}
 (d_{AR}^{\dagger} d_{BL} + d_{BR}^{\dagger} d_{AL})v_-^{10}
 (Y^{10}_{AB}+Y^{10}_{BA})
\label{33} 
\end{equation}
and the hermitian conjugate adds
\begin{equation}
 (d_{BL}^{\dagger} d_{AR} + d_{AL}^{\dagger} d_{BR})
  [v_-^{10} (Y^{10}_{AB}+Y^{10}_{BA})]\sp * 
\label{34}
\end{equation}
 Obviously $Y_{AB}=Y_{BA}$ follows without loss of generality. 
 These two equations can be combined and writen in terms of 
 4-component Dirac spinors, $d_A$ and $d_B$ as
 \begin{equation}
 2[Re(v_-^{10} Y^{10}_{AB})\bar{d}_{A} d_B +
 i Im(v_-^{10} Y^{10}_{AB})\bar{d}_{A} \gamma_5 d_{B}],
\label{35}
\end{equation}
\begin{equation}
 2[Re(v_-^{10} Y^{10}_{AB})\bar{d}_{B} d_A +
 i Im(v_-^{10} Y^{10}_{AB})\bar{d}_{B} \gamma_5 d_{A}].
\label{36}
\end{equation}
 and if $v_-^{10} Y^{10}_{AB}$ is real the pseudoscalar term vanishes
 and there is a real Dirac mass contribution 
 to a symmetric flavour mass matrix. The other off-diagonal 
  elements are of course completely similar and the diagonal 
 contributions are real and scalar if $v_-^{10}Y^{10}_{AA}$ are real.   
 Thus the flavour mass matrix is Hermitian. Also the mass contribution 
 resulting conserves CP as well as P. 

 \subsection{Three $\Gamma$}
 These give the $\{120\}$ anti-symmetric representations which are 
 developed here as imaginary hermitian matrix representations.
 
 (i) As shown in the Table there are two types of colour 
singlet neutral vevs  associated with three $\Gamma 's$.
 In what follows the evaluation  for the type

  {\bf $\Gamma_7\Gamma_8\chi_5$,
 $\Gamma_7\Gamma_8\chi_5^{\dagger}$}

 is outlined. The flavour mass matrix elements for the down 
 quarks are included in
 \begin{eqnarray}
 [\left< \psi_{+A}^{\star} \right| B_{\Gamma}C^{-1}
 \Gamma_7\Gamma_8\chi_5 \left|\psi_{+B}\right>Y^{120}_{AB}+\nonumber\\
 \left< \psi_{+B}^{\star} \right| B_{\Gamma}C^{-1}
 \Gamma_7\Gamma_8\chi_5 \left|\psi_{+A}\right>Y^{120}_{BA}]v_-^{120a}
  +h.c.
\label{37}
 \end{eqnarray}   
 corresponding to equation (\ref{32}). 
 The total result of the calculation is
\begin{eqnarray}
 -i[(d_{AR}^{\dagger} d_{BL} - d_{BR}^{\dagger} d_{AL})]
 v_-^{120a} (Y^{120}_{AB}-Y^{120}_{BA})\nonumber\\
 +i[(d_{BL}^{\dagger} d_{AR} - d_{AL}^{\dagger} d_{BR})]
 [v_-^{120a}(Y^{120}_{AB}-Y^{120}_{BA})]^* 
\label{38}
\end{eqnarray}

Obviously $Y^{120}_{AB}=-Y^{120}_{BA}$ follows without loss of 
 generality.  These two lines can be rearranged and written
 in terms of  4-component Dirac spinors, $d_A$ and $d_B$ as
 \begin{equation}
 -2i[Re(v_-^{120a} Y^{120}_{AB})\bar{d}_{A} d_B +
 i Im(v_-^{120a} Y^{120}_{AB})\bar{d}_{A} \gamma_5 d_{B}],
\label{39}
\end{equation}
\begin{equation}
 +2i[Re(v_-^{120a} Y^{120}_{AB})\bar{d}_{B} d_A +
 i Im(v_-^{120a} Y^{120}_{AB})\bar{d}_{B} \gamma_5 d_{A}]].
\label{310}
\end{equation}

 Suppose $v_-^{120a}Y_{AB}$ to be real. Then there is only a scalar 
 mass term, the same applying to all the other off-diagonal
 contributions. The diagonal terms are anyway zero and thus
 the total contribution is imaginary anti-symmetric,so of 
 hermitian flavour matrix form.  

 The extra factor $i$ with a $3\Gamma$ operator such as 
 $\Gamma_7\Gamma_8\chi_5$ is because $\Gamma_7\Gamma_8=
 i(\chi_4^{\dagger}\chi_4-\chi_4\chi_4^{\dagger})$.
 Thus it might be thought that the association of
 hermiticity with a purely scalar mass term is due to a 
 particular choice of phase in $\Gamma_7\Gamma_8\chi_5$.
 This is not so; multiplication of that operator by
 $e^{i\phi}$ still yields the same association  This can be  shown
 by explicit calculation but generally one can reason as follows. 

The mass term in the Lagrangian of 3-flavoured chiral theories is
\begin{equation}
    \psi_R^{\dagger}M\psi_L+\psi_L^{\dagger}M^{\dagger}\psi_R.
\label{311}
\end{equation}
where $M$ is a $3 \times 3$ flavour matrix and $\psi_R,\psi_L$ are 
2-component spinors with the three flavour index implicit. For 
Hermitian matrices, $M^{\dagger}=M$, this converts trivially into
4-component spinor, $\psi$, as a scalar, $\bar {\psi} M \psi$,
 and upon flavour diagonalization results in normal Dirac equations.
 However if $M$ is not Hermitian
 \begin{equation}
    \psi_R^{\dagger}M\psi_L+\psi_L^{\dagger}M^{\dagger}\psi_R.=
{1 \over 2}\bar {\psi} (M+M^{\dagger}) \psi -
{1 \over 2}\bar {\psi} \gamma_5 (M-M^{\dagger}) \psi
\label{312}
\end{equation}
and the pseudoscalar parity breaking $\gamma_5$ term intrudes upon
 the canonical Dirac equation. Conversely the presence of a $\gamma_5$
 term breaks hermiticity.

 (ii) Using the notation 
 $ \Gamma\Gamma \equiv 
 \Gamma_1\Gamma_2 +\Gamma_3\Gamma_4 +\Gamma_5\Gamma_6 $ 
 it is seen from the Table that the other type of 
 $3\Gamma$ operator is

 {\bf $\Gamma\Gamma\chi_5,  \Gamma\Gamma\chi_5^{\dagger}$}. 

 For the down quarks this again yields the expressions
 (\ref{38})-(\ref{310}) with $v_-^{120b}$ instead of
 $v_-^{120a}$.  Thus the association of hermiticity with
 a purely scalar mass term holds here too; and it 
 equally applies for the masses of all particles.

 For three $\Gamma$ in both(i) and (ii) above and contrary to the 
 cases of one $\Gamma$  and five $\Gamma$
 the adopted purely scalar case violates CP because of the
 anti-symmetry of the $\{120\}$.

\subsection{Five $\Gamma$}
 
 These arise from the $\{126\}$.
 As shown in the Table mass terms are generated by

 {\bf  $\Gamma_7\Gamma_8\Gamma\Gamma\chi_5$} and 
 {\bf $\Gamma_7\Gamma_8\Gamma\Gamma\chi_5^{\dagger}$}

 These indeed give results like the one $\Gamma$ case in the 
 respect that if $v_-^{126a} Y^{126}_{AB}$ is real the pseudoscalar 
 term vanishes and there is a real scalar mass contribution 
 to a symmetric flavour mass matrix. That is a scalar 
 mass matrix contribution implies that it is hermitian, 
 and vice versa. As before the same applies to all the 
 particle masses given by these operators. As in the 
 One $\Gamma$ case the resulting contribution also conserves CP.

\subsection{summary}

 Collecting now the contributions from just the first eight lines 
 of the Table these also include the familiar operator coupling
 right chiral neutrinos to left chiral neutrinos which is here denoted
 by $M^n$. All these together result in the mass operators with
 obvious labels 

\begin{equation}
M^u_{AB}=
 2Y^{10}_{AB}v^{10}_+ +2Y^{126}_{AB}v^{126a}_+
 +2iY^{120}_{AB}v^{120a}_+ -2iY^{120}_{AB}v^{120b}_+
\label{313}
\end{equation}
\begin{equation}
M^n_{AB}=
 2Y^{10}_{AB}v^{10}_+ -6Y^{126}_{AB}v^{126a}_+
 +2iY^{120}_{AB}v^{120a}_+ +6iY^{120}_{AB}v^{120b}_+
\label{314}
\end{equation}
\begin{equation}
M^d_{AB}=
 2Y^{10}_{AB}v^{10}_- -2Y^{126}_{AB}v^{126a}_-
 -2iY^{120}_{AB}v^{120a}_- -2iY^{120}_{AB}v^{120b}_-
\label{315}
\end{equation}
\begin{equation}
 M^e_{AB}=
 2Y^{10}_{AB}v^{10}_+ +6Y^{126}_{AB}v^{126a}_-
 -2iY^{120}_{AB}v^{120a}_- +6iY^{120}_{AB}v^{120b}_-
\label{316}
\end{equation}

where the Yukawas $Y^{10}_{AB}$ and $Y^{126}_{AB}$ are real and 
symmetric while the $Y^{120}_{AB}$ are real and anti-symmetric.
 The vacuum expectation values $v$ are also real to make all the 
 above mass matrices hermitian.
 
 It should be noted that the proof of hermiticity only evidently
 carries through in the fermion basis implicit in the present model;
 chiral rotations different for different flavours can change the
 coefficients of the scalar and pseudoscalar bilinears, changing the
 flavour mass matrices.

 The last two 5$\Gamma$ lines  of the Table do not contribute
 mass to the quarks or charged leptons, but only to the neutrinos.
 This arises from evaluation of equations (\ref{23}) using 
 (\ref{21}) and (\ref{22}) and the associated particle assignments therein.
 These are given in the Appendix and associated with the left-right
 neutrinos of eqn(\ref{314}). First order block diagonalization of the
 resulting $6 \times 6$ flavour mass matrix gives 3 light Majorana
 neutrinos and 3 heavy. 

\section{Mass Relations}

 The 16-plet of Weyl fermions eqns.(\ref{21},\ref{22}) has chiral
 neutrinos, $n_L$, $(n_R)^c$.We denote the 4-component neutrinos
 corresponding to $n_L$, $n_R$ as $\nu_L$, $N_R$ respectively
 (see the Appendix for further details). The
 see-saw hypothesis assigns a large mass to $N_R$ through a
 Majorana mass term, with flavour matrix (arising from eqn.(\ref{23})
 and the last line of the table) here denoted $M$.
 These left and right neutrinos couple together with flavour
 matrix $M^n$, given in eq.(\ref{314}), analogous to the 
 quark and charged lepton mass matrices. The penultimate
 line of the Table gives rise to a Majorana mass term for
 $\nu_L$ The flavour matrix for this, numerically very small
 compared to $M$, is denoted by $m$. Diagonalization  
 of the resulting matrix
\begin{equation}
M^{neutrinos}=
\left[\begin{array}{cc}
m & M^n \\
M^{nT} & M 
\end{array}\right]
\label{41}
\end{equation}
results in 3 light physical Majorana neutrinos and 
3 heavy physical neutrinos in the top left and bottom right
 respectively of $M^{neutrinos}$. The light neutrino Lagrangian
 has: (i) a contribution having P and CP invariance containing
 a $3 \times 3$ Hermitian flavour mass matrix; (ii) a contribution
 violating both P and CP containing a $3 \times 3$ imaginary 
 non-Hermitian flavour mass matrix. This latter term arises from 
 the original ('Type I') seesaw mechanism and is the only P violating
 term at  high energy in the resulting mass Lagrangian of the model. 

 For analyzing, as now follows, 
 the mass relations of the quarks and charged 
 leptons the notation is simplified, dropping the flavour indices 
 from the Yukawa matrices, Y, and defining new matrices which 
 incorporate the scalar vacuum expectation values  
 (such as $v^{10}_{\pm}$ for those in the \{10\}).

\begin{equation}
 h=Y^{10}v^{10}_-, f=Y^{126a}v^{126}_-, g=Y^{120}v^{120a}_-.     
\label{42}
\end{equation}
 Certain ratios of the real vacuum expectation values, 
 required for the mass equations (\ref{45}) to (\ref{48})
 below, are 
 
 \begin{eqnarray}
 r_h=v_+^{10}/v_-^{10},r_f=v_+^{126a}/v_-^{126a},
 r_g=v_+^{120a}/v_-^{120a},\nonumber\\
r_1=v_-^{120b}/v_-^{120a},
 r_2=v_+^{120b}/v_+^{120a}.
\label{43}     
\end{eqnarray}
 In addition for the Majorana neutrino matrices
 $M=r_Mf$ and $m=r_mf$ 
\begin{equation}
  r_M=v_+^{126b}/v_-^{126a},r_m=v_-^{126b}/v_-^{126a}
\label{44}
\end{equation}
With these notations the quark and lepton mass matrix equations
 are
\begin{eqnarray}
M^d = h-f-ig(1+r_1) \label{45}\\
M^e = h+3f-ig(1-3r_1) \label{46}\\
M^u = r_h h+r_f f+ir_g g(1-r_2) \label{47}\\
M^{n} = r_h h-3r_f f+ir_g g(1+3r_2) \label{48}
\end{eqnarray}
where the mass matrices are hermitian with $h,f$ being real 
symmetric and $g$  real antisymetric. As noted in the
 introduction the hermiticity is a significant difference 
 from the majority of previous papers. (Allowing for changes
 due to conventions the part of these equations in $h$ and $f$
 are recognisably the same as those written in very many previous
 papers such as references \cite{GMN} - \cite{GK1}. On the other
 hand the terms in $g$ are formally, and physically, different
 as they involve the ratios $r_1$ and $r_2$ arising from the
 equations of section III.D.)

 Putting aside neutrino masses and mixing to be considered 
 later, the present data, some of it being significantly only 
 approximate, is 9 quark and charged lepton masses,
 3 CKM matrix angles and 1 phase.

 To make use of the CKM matrix ($V=U_d^{\dagger}U_u$) 
data it is a common device,when possible, to take a 
basis in which either the d-quark matrix is real diagonal
 (implying $U_d$ is unity) or the u-quark matrix is
 (so $U_u$ is unity). Then either $U_u$ or $U_d^{\dagger}$
 respectively is the CKM matrix. While this might be done
 in the general case of the present model the unitary 
 matrix required to change the basis generally bestows 
 imaginary parts on the real 
 matrices $h,f,g$, thus upsetting a simplifying 
 feature of the model. However we can avoid this upset
 and shall make use of 
 this device by considering some special cases.

\subsection{Special Cases}
 To retain those features and the (relative) simplicity 
 of numerical calculations there are two special cases
 of the vevs associated with the two couplings (which we
  have denoted as 120a and 120b respectively) in the
 $\{120\}$. These are 
 (i) $v_+^{120b}=v_+^{120a} \Rightarrow r_2=1$, making 
 $M^u$ real symmetric, diagonalisable by a real 
 orthogonal change of basis; and
 (ii) $v_-^{120b}=-v_-^{120a} \Rightarrow r_1=-1$, making 
 $M^d$ real symmetric, diagonalisable by a real 
 orthogonal change of basis. The latter special case
 involves importing a relative phase of $\pi$ which 
 however preserves the reality conditions of the model.
 Taking either of these special cases, with their 
 associated change of basis, preserves the real symmetric
 (anti-symmetric) nature of the matrices $h,f,(g)$. 
  
 To illustrate the numerical evaluation consider the 
 special case (ii) which will be used in the following
 section on neutrino masses and mixing. The mass matrix
 equations simplify to 
\begin{eqnarray}
M^d = h-f \label{410}\\
M^e = h+3f-4ig \label{411}\\
M^u = r_h h+r_f f+ir_g g(1-r_2) \label{412}\\
M^{n} = r_h h-3r_f f+ir_g g(1+3r_2) \label{413}
\end{eqnarray} 
 The first three equations yield the mass relation
\begin{equation}\label{414}
 M^e = x M^d + y Re(M^u) + izIm(M^u)    
\end{equation} 
which,equating matrix coefficients in the real and 
 imaginary parts of equations (\ref{410}),(\ref{411}),
(\ref{412}), has 
\begin{eqnarray}
x=(r_f-3r_h)/(r_h + r_f),\nonumber\\
y=4/(r_h + r_f),\nonumber\\
z=-4/r_g(1- r_2). \label{415}
\end{eqnarray}

The three matrices on the right hand side of equation(\ref{414})
are  evaluated as follows in terms of the quark masses called 
here $d,s,b$ and $u,c,t$,and the CKM matrix $V$. Now 
$M^d,Re(M^u),Im(M^u)$ are written respectively as

$$
\left[\begin{array}{ccc}
d & 0 & 0\\
0 & s & 0\\
0 & 0 & b
\end{array}\right],
\left[\begin{array}{ccc}
w_1 & u_1 & u_3\\
u_1 & w_2 & u_2\\
u_3 & u_2 & w_3
\end{array}\right],
\left[\begin{array}{ccc}
  0  & v_1  & v_3\\
-v_1 & 0    & v_2\\
-v_3 & -v_2 &   0
\end{array}\right].
$$
 As stated above since the basis is diagonal in the down quarks
and all the matrices are Hermitian then $M^u$ is diagonalized 
by the CKM matrix, $V$, giving the real parameters $u_i,v_i,w_i$
in terms of $u,c,t$ and the CKM matrix by
\begin{equation}
 M^u = V^{\dagger} M^u_{diag} V
\label{416}     
\end{equation}
 the diagonal matrix 
$M^u_{diag}$ having elements $u,c,t$.

The choice of the CKM matrix is the additional input at this stage.

Given a matrix, (i) its trace,(ii)the trace of the inverse and (iii)
 the determinant are each invariant under a unitary transformation.
 This leads to the known way of solving equations such as(\ref{414}) 
since a unitary transformation can transform the Hermitian $M^e$
into a diagonal matrix  of the charged lepton masses. Then we equate
 the three invariants of the two 
 sides of equation (\ref{414}). For the right hand side (i) gives an 
expression linear in $x,y$ while (ii) being equivalent to the sum of
the $2 \times 2$ diagonal sub-matrices gives a quadratic in $(x, y,z)$
but with $z$ only occuring as $z^2$ and (iii) yields a cubic but again
with $z$ only occuring as $z^2$. 

The coefficients are real and elimination gives a cubic equation,
 either in $x$ or $y$, which can be solved by an analytic expression
and on evaluation yields numbers for the set $ {x,y,z^2}$. Only those 
sets with $x,y$ real and $z^2>0$ are acceptable solutions of
 equation(\ref{414}). 

\section{Neutrino Masses and Mixings}

Continuing with the special case where $r_1=-1$ the light neutrino 
 masses and mixing are due for consideration. The approach is first 
 to identify the three remaining free real parameters of the model. 
 Then to choose a possible set of
 3 neutrino masses which are in accord with mass data and search 
 for a solution of those parameters (in principle there could be 
 more than one) for which the model gives the chosen masses. 
 This can be repeated for various choices of light neutrino masses.
  For each solution set the model is fixed with all parameters
 determined by the assumed quark, charged lepton and
 light neutrino masses and CKM matrix.
 Each parameter-fixed model predicts an MNS mixing matrix which
 can be considered for plausibility.

 So far use has been made of 13 assumed data points - 9 quark and 
 charged lepton masses, 3 CKM angles and 1 phase - albeit some of
 these are subject to considerable error. Using  
 equations (\ref{410}),(\ref{411}),(\ref{412}),(\ref{416})
 and a basis diagonal in the mass matrix $M^d$ the hermitian 
 quark and charged lepton mass matrices $M^d$,$M^u$,$M^e$
 have been synthesized from the 13 data points. An alternative
 description, in terms of the quantities on the RHS of
 equations (\ref{410}),(\ref{411}),(\ref{412}),(\ref{413}) is 
 that the real matrices $h,f,g$ along with the real parameters
 $r_h,r_f,r_g(1-r_2)$ have been constructed.(There seem to be 
 15 constructed parameters from 13 data points. However these 15 are 
 not independent because eq.(\ref{416}), arising from the special
 assumptions, produces 9 parameters from 7 data.)

 Turning now to the neutrinos, a little manipulation of the equations
 yields the neutrino mass matrix of eq.(\ref{413}) as
 \begin{equation}
 M^{n}={1-x \over y}M^d - {2+x \over y}4\tilde M +
 i{1+3r_2 \over 1-r_2}Im(M^u)     
\label{51}
\end{equation}
 \begin{equation}
 \tilde M \equiv (Re(M^e)-M^d)/4 =f=M/r_M=m/r_m.     
\label{52}
\end{equation}
 In eq.(\ref{51}) $r_2$ occurs. It is a free parameter since 
 only $r_g(1-r_2)$ has been constructed.

 The 'seesaw' mass matrices $m$ and $M$ of eq. (\ref{41}) 
 (arising from the $\{126\}$ representations of the last two
 rows of the Table) are proportional to the matrix $f$ as 
 displayed in (\ref{52}). Thus there are 2 more free parameters,
 $r_m$ and $r_M$, in addition to $r_2$, leaving the model with
 just 3 so far undetermined real (and dimensionless)
 parameters.These can in priciple
 be fixed by fitting 3 given light neutrino masses. Since data is 
 only known on mass differences exploration requires postulation 
 of one light neutrino mass.

 By a redefinition of the neutrino states, working to first order 
 in the numerically small matrix $\eta = M^{n}M^{-1}$, 
 as shown in the Appendix, $M^{neutrinos}$ can be transformed to
 block diagonal form
$$
\left[\begin{array}{cc}
m_{\nu} & 0 \\
0 & M 
\end{array}\right].
$$
 The $3 \times 3$ matrix of the small
 mass neutrinos is given by the usual seesaw formula
 (Type II and Type I) $m_{\nu}=m-M^nM^{-1}M^{nT}$. For the present
 special case the matrix $m=r_m f$, contributed wholly by the 
 $\{126\}$, is hermitian, being real and symmetric. The Type I
 contribution, $-M^nM^{-1}M^{nT}$, though composed from hermitian
 matrices is neither real nor hermitian. However it is symmetric thus
 $m_{\nu}$can be written in terms of real symmetric matrices $m_1,m_2$  
 as $m_{\nu} = m_1 + im_2$. The eigenvalues of the hermitian matrix  
 \begin{equation}
 m_{\nu}m_{\nu}^{\dagger} = (m_1 + im_2)(m_1 - im_2)
\label{53}     
\end{equation} 
 are the neutrino masses squared.

The questions now are: (i) are the 3 free parameters sufficient to
fit some possible sets of experimental neutrino masses; (ii) if so 
is agreement also found with our knowledge of the MNS matrix

\subsection{An Example}

 The Yukawa matrices,$Y^{10},Y^{126},Y^{120}$ have been assumed
 throughout to be appropriated to the ${10}$,${126}$,${120}$
 SO(10) representations respectively.
 While this paper is not committed to a GUT model nevertheless 
 the quark and lepton masses used should be those appropriate to 
 a high energy where parity and SO(10) are restored, thus involving
 extrapolation by renormalization group equations.
 There are well known (and well used)extrapolations
 by Das and Parida \cite{DP} including renormalization group equations
 of the standard (nonSUSY) model, the 2 Higgs doublet model and the 
 minimum supersymmetric model. These have the general feature of such
 extrapolations of shifting the quark masses by considerably more than
 the relative shift of the lepton masses. For the light neutrino masses 
 the present low energy data is here used as a guide.
   
 Possible sets of extrapolated quark and charged lepton masses and CKM
 matrix using the results of Das and Parida \cite{DP} have been
 constructed and used in various papers. We make use of some of those 
 previous works by adopting the data sets (appropriate to an energy
 of $2 \times 10^{16}$ GeV of (i) Goh et al. \cite{GMN} and (ii)
 Bertolini et al.\cite{Bertolini}). However it is not at all
  the purpose in this paper to make an assiduous search for 
 solutions plausible on some criteria. Rather it is to sample
 sparsely to illustrate some possibilties albeit in a special
 case ($r_1=-1$) of the original model.

  For the present experimental data the  
 3-neutrino mixing scheme reviewed by B. Kayser \cite{Kayser}
 is used. The neutrinos being named as 1,2,3 the following central
 values of the difference of squared masses are adopted
 $$(\Delta m^2)_{21}=8.0\times 10^{-5} eV^2,
 (\Delta m^2)_{32}=2.5\times 10^{-3} eV^2, $$ along with the assumption 
 of a neutrino hierarchy $m_{\nu 1}^2 < m_{\nu 2}^2 < m_{\nu 3}^2$ so 
 that  
 \begin{eqnarray}
m_{\nu 2}^2=m_{\nu 1}^2+(\Delta m^2)_{21}\label{54}\\
m_{\nu 3}^2=m_{\nu 2}^2+(\Delta m^2)_{32}. \label{55}
\end{eqnarray} 
 The computing code calculates theoretical values of these 3 neutrino
 masses squared: $\mu_i^2(r_2,r_m,r_M)$, each of these three being 
 functions of the free parameters $r_2,r_m,r_M$. It is necessary to
 find values of $(r_2,r_m,r_M)$ so that the $\mu_i^2$ are equal
 to the $m_{\nu i}^2$, $(i=1,2,3)$. This is done in an obvious way
 by inventing and optimizing a function whose extreme value
 (say zero) is reached
 when  $\mu_i^2=m_{\nu i}^2$ for all $(i=1,2,3)$, thus achieving an
 exact solution. The principle of this procedure is not the same as 
 finding the 'best fit' (for example by $\chi^2$ minimization) of  
 many  parameters. Because of the complication of the numerical
 calculation of the neutrino masses it is almost necessary to use a 
 non-derivative method. So a non-derivative simplex method \cite{NR},
 operating in the three-dimensional space of $(r_2,r_m,r_M)$ is used 
 to find any zeros of the chosen function. The computer search is
 specialised to those regions of small $r_m$ and large $r_M$ 
 suitable to produce small neutrino masses.  

 (i) The paper of Goh et al.\cite{GMN} uses the Das and Parida
 extrapolation of masses
  to $2 \times 10^{16}$ GeV by the MSSM RGE ($tan(\beta)=10$ in
 Table II of \cite{DP}) and also quotes the real part of the CKM m
 matrix. These are used as data input in the present example. Using
 the Wolfenstein parametrization gives Im(CKM) as
\begin{equation}
 \left[\begin{array}{ccc}
    0   &    0    & -.00320\\
    0   & -.00059 &  .00074\\
-.00328 &    0    &     0
\end{array}\right], 
 \label{ImCKM}
 \end{equation}

 In all this $m_{\nu 1}^2$ is a quantity of choice being part of any 
 postulate on neutrino masses. A first idea is that it should be small
 but not negligable, say of the order of $1.0\times 10^{-5} eV^2$. In 
 this region the optimization code gives the desired equality to very
 great accuracy. For $m_{\nu 1}^2=2.0\times 10^{-5} eV^2$,
 $m_{\nu 2}^2=1.0\times 10^{-4} eV^2$,
 $m_{\nu 3}^2=2.6\times 10^{-3} eV^2$ then then the dimensionless
 free parameters of equations (\ref{51}) and (\ref{52}) have the values   
 \begin{equation}
  r_2 = 3.13 ,r_m= 2.1 \times 10^{-9},r_M =5.9 \times 10^{15}. 
\label{56}
\end{equation} 

 Having fixed the free parameters, the only extra input having been the
 partly hypothesized neutrino masses, it is then possible to 
 calculate the MSN matrix. Rather surprisingly one finds good
 agreement with  presently accepted features of ths matrix.

 The charged lepton mass matrix, $M^e$, (now calculable) is hermitian
 in the model and so diagonalizable by a unitary transformation
 \begin{equation}
  X_e^{\dagger} M^e X_e = M^e_{diag} 
\label{57}
\end{equation} 
 The light neutrino mass matrix, $m_{\nu} = m_1 + im_2$, and the 
 matrix for the masses squared $(m_1 + im_2)(m_1 - im_2)$ is hermitian
and can be diagonalized by a unitary transformation as
 \begin{equation}
  X_{\nu}^{\dagger} (m_1 + im_2)(m_1 - im_2) X_{\nu}.  
\label{58}
\end{equation}
 The diagonalization of $m_{\nu}$ requires a bi-unitary transformation
 as 
  \begin{equation}
  X_{\nu}^{\dagger} (m_1 + im_2) Y_{\nu}.  
\label{59}
\end{equation}
 The MNS matrix, similarly to the CKM matrix, is defined as 
 \begin{equation}
   U_{MNS} = X_{e}^{\dagger} X_{\nu}.  
\label{510}
\end{equation}
 and $U$ can be calculated in the present model using the values
 of the 3 free parameters fixed by using values of the neutrino
 masses as discussed above. In this particular case the result
 for the matrix of moduli squared of the elements of U
 ($\left\vert U_{e\nu} \right\vert^2$) is
\begin{equation}
 \left[\begin{array}{ccc}
.638 & .344 & .017\\
.260 & .331 & .409\\
.102 & .325 & .573
\end{array}\right], 
 \label{511}
 \end{equation}
 bearing a distinct resemblance to the postulated 'ideal' structure 
 of this matrix in tri-bimaximal mixing \cite{HPS}:
\begin{equation}
 \left[\begin{array}{ccc}
2/3 & 1/3 & 0\\
1/6 & 1/3 & 1/2\\
1/6 & 1/3 & 1/2
\end{array}\right], 
 \label{512}
 \end{equation}

 It may be noted that the 
13, 23 and 12 elements are clearly within the range of many 
analyses of the experiments \cite{HPS}. (It is interesting 
that in this solution the mass matrix elements contribution
 from Type II are much greater than those from Type I except 
 for the (3,3) where the contributions are of the same order
 of magnitude.) Similar results hold for all values of 
 $m_{\nu 1}^2$ between about $1.6 \times 10^{-5}$ and
 $2.4 \times 10^{-5}$ in $eV^2$.

 (ii) Bertolini et al. \cite{Bertolini},making some use of
 the Das and Parida \cite{DP} extrapolation of masses to
 $2 \times 10^{16}$ GeV by the MSSM RGE ($tan(\beta)=10$)
 have given quark and charged lepton masses and 3 CKM angles
 plus the phase angle. Bertolini et al. \cite{Bertolini}
 have revised the central 
 values of the extrapolated masses of the lightest quarks
 to $0.55,1.24,21.7$ MeV whereas Goh et al.\cite{GMN}, as
 used in (i), have $0.72,1.5,30.0$ MeV. The extrapolated
 masses of the three heaviest quarks and the charged leptons
 remain the same as those in  \cite{GMN}.Some off-diagonal
 elements of the CKM matrix also display some non-trivial
 differences from those of Goh et al \cite{GMN} used in (i).

 In the fitting, of the theory to the pseudo-physical 
 neutrino masses, some samples of $m_{\nu 1}^2$ were taken 
 in and around the same region as in (i) above. Any that 
 succeeded in fitting yielded significantly different 
 parameters  from those in (i) resulting in spectacularly
 implausible MSN matrices. Further numerical investigation
 seemed to show that each one of the data changes mentioned
 above had influence on the results. That is results are
 sensitive to changes in the least well known data.
 It should be emphasized that neither in case (i) nor case (ii)
 was there attempted extensive investigation of very many
 neutrino spectra.

 \section{Summary}
 The model makes use of SO(10) but with multiplets restricted
 in number and kind. 

 Firstly the philosophy is to adopt the 16-plet of fermions 
 as the particles we know and only consider couplings arising 
 from ${16} \times {16}= {10}+{120}+\overline{126}$. Secondly to restrict
 those couplings so that the resulting mass matrices are 
 hermitian at high energy; an argument is given that within SO(10) this
 can be formulated naturally; also the restoration at high energy of 
 parity invariance of Lagrangian mass terms implies hermiticity.
 The scalar vacuum expection values are 
 classified in the Pati-Salam subgroup of SO(10) using the 
 Clifford algebra representation; this distinguishes two 
 realisations of the $\{120\}$ associated with different vacuum 
 expection values. Assuming the values for the quark and
 charged lepton  masses and the CKM angles and phase leaves
 just 4 undetermined real dimensionless parameters in the theory.
 Fixing one  of these parameters to a special value, thus
 simplifying the calculation and the range of results, leaves just
 3 free parameters. To fix these the hierarchical hypothesis
 for light neutrino masses is adopted, together with specifying 
 the lowest neutrino mass (at various values) to give the 3 light
 neutrino masses in accord with existing data on neutrino mass squared
 differences. The theory then predicts the MSN matrix. In the
 case of one set of quark and charged lepton masses and CKM
 matrix \cite{GMN},and a range of lowest neutrino masses, 
 the MSN matrix is, rather surprisingly, in accord with the
 tri-bimaximal mixing suggested by the data \cite{HPS}. This
 result does does not hold for another data set \cite{Bertolini},
 for the same physical quantities, that was tried. Solutions thus 
 appear sensitive to changes in the less well known physical 
 quantities.   

 {\bf Acknowledgments.} I thank David Sutherland for many valuable
 discussions and Colin Froggatt for comments on the manuscript.

\newpage

{\bf APPENDIX}

\appendix

In addition to the neutrino content of the standard model which 
has just left chiral (and massless) neutrinos, the 16-plet of SO(10)
 has both left chiral, $n_L$, and right chiral, $n_R$,
  neutrinos. These occur in the 16-plet vector (\ref{22})
  as $\tilde \psi_5$ and $-\psi_o^c$ respectively, being 
 2-component Weyl spinors.. 

 The matrix elements (\ref{23b}) give rise to three  
 mass matrices for the neutrinos. Two of these are the self 
 couplings of $n_L$ and $n_R$ respectively. Both of these are 
 Majorana mass terms. The third couples $n_L$ and $n_R$ and
 gives rise to the mass term $M^n$which appears in equations
 (\ref{48}) and (\ref{413}). The terms are as follows.

 (i) The self couplings of $n_R$ arise from
 $\chi_1^{\dagger}\chi_2^{\dagger}
\chi_3^{\dagger}\chi_4^{\dagger}\chi_5^{\dagger}$ as in 
the last line of the table. They are
$$(n_{AR})^{cT} C^{-1}(n_{BR})^{cT}v_+^{126b}Y_{AB}^{126}+hc$$ 
where $A,B$ are generation indices. These can be put into
4 component spinor notation by defining  4 component right chiral 
spinors $$ N_{AR}^T \equiv (0,n_{AR})$$ giving the (Majorana) mass 
terms as $$ {1 \over 2}\overline{N_{AR}^c}N_{BR} M_{AB} +hc$$
\begin{equation}
 M_{AB}=2 v_+^{126b}Y_{AB}^{126}
\label{A13}
\end{equation}
 Since $v_+^{126b}$ and $Y_{AB}^{126}$ are real $M$ is a real
symmetric matrix.

 (ii) The self  couplings of $n_L$ arise from 
 $X=\chi_1\chi_2\chi_3\chi_4\chi_5^{\dagger}$ as in the
 penultimate line of the Table. Analogously to (i) above the 
 resulting 2-component (Weyl)spinor results can be expressed
 in 4 component spinor terms by defining 
 $$ \nu_{AL}^T \equiv (n_{AL},0)$$ giving the (Majorana) mass 
 as $${1 \over 2} \overline{\nu_{AL}^c}\nu_{BL}m_{AB} +hc$$.
 \begin{equation}
 m_{AB}=2v_-^{126b}Y_{AB}^{126}
\label{A14}
\end{equation}
 Since $v_-^{126b}$ and $Y_{AB}^{126}$ are real $m$ is a real
symmetric matrix.

 (iii) The couplings of $n_R$ to $n_L$ arise from
 $X$ as in the first 8 lines of the Table.In 4-component 
 spinors these give mass terms
 $$\overline{\nu_{AL}}N_{BR}M^n_{AB} +hc$$   
 where $M^n$ is the hermitian mass matrix of equation
 (\ref{48}).

 These neutrino mass terms can be written in matrix form,
 using $\overline{\nu_L}M^nN_R=\overline{N_R^c}M^{nT}\nu_L^c$
 \cite{Buch}, as
$$
{1 \over 2}
\left[\begin{array}{cc}
 \overline{\nu_L} & \overline{N_R^c} 
\end{array}\right]
\left[\begin{array}{cc}
m & M^n \\
M^{nT} & M 
\end{array}\right]
\left[\begin{array}{c}
\nu_L^c \\
N_R 
\end{array}\right]
+ hc.
$$

The matrix $M$ is very large compared to $M^n$ and $m$.
So, working to first order in $M^{-1}$, the seesaw method 
can be implemented by defining  
 \begin{equation}
 L=\nu_L + (M^{-1}m)_R^cN,R^c=N_R+(M^{-1T}m)\nu_L
\label{A15}
\end{equation}
 noting that $M$ and $m$ are hermitian matrices. The neutrino
 mass terms become
$$
{1 \over 2}
\left[\begin{array}{cc}
 \overline{L} & \overline{R^c} 
\end{array}\right]
\left[\begin{array}{cc}
m_{\nu} & 0 \\
0 & M 
\end{array}\right]
\left[\begin{array}{c}
L^c \\
R 
\end{array}\right]
+ hc.
$$
 \begin{equation}
 m_{\nu}=m+m_{\nu}^I
\label{A16}
\end{equation}
 \begin{equation}
 m_{\nu}^I=-M^nM^{-1}M^{nT}
\label{A17}
\end{equation}

 Eqn(\ref{A17}) gives the original seesaw term, often known as Type I
  seesaw. $m$ is hermitian and $m_{\nu}^I$ is complex symmetric.Denoted
 $(m_{\nu},M)$ are the flavour mass matrices of the (light,heavy) 
 Majorana neutrinos with Majorana fields \cite{Buch} 
\begin{equation}
 \nu=L+L^c , N=R^c+R.     
\label{A18)}
\end{equation}
with the mass terms for the light neutrinos being
 \begin{eqnarray}
 {1 \over 2}\overline{\nu} m_{\nu} (1+\gamma_5) \nu  +hc\nonumber\\
 = \overline{\nu} m \nu
 + \overline{\nu}[Re( m_{\nu})+i\gamma_5 Im( m_{\nu})] \nu.
\label{A19}     
\end{eqnarray}

 In eqn(\ref{A19}) the two first terms, containing hermitian matrices,
 are P and CP invariant but the $i\gamma_5$ symmetric matrix 
 term violates CP as well as P. Thus P and CP violation in the light
 Majorana neutrino mass terms arise solely from the original seesaw
 mechanism involving the non-hermitian product of 3 hermitian matrices.

\end{document}